\documentclass[twocolumn]{revtex4}
\usepackage[utf8]{inputenc}
\usepackage{graphicx}
\usepackage{color}

\begin{document}

\title{Keffer-like form of the symmetric Heisenberg exchange integral:
Contribution to the Landau--Lifshitz--Gilbert equation and spin wave dispersion dependence}

\author{Pavel A. Andreev}
\email{andreevpa@my.msu.ru}
\affiliation{Department of General Physics, Faculty of physics, Lomonosov Moscow State University, Moscow, Russian Federation, 119991.}

%andreevpa@physics.msu.ru

%\date{\today}

\begin{abstract}
The symmetric Heisenberg exchange interaction and antisymmetric Dzyaloshinskii-Moriya interaction
are parts of the tensor potential describing effective spin-spin interaction caused by the superexchange interaction of magnetic ions via nonmagnetic ion.
There is the Keffer form of the vector constant of the Dzyaloshinskii-Moriya interaction,
which includes the shift of the nonmagnetic ion (ligand) from the line connecting two magnetic ions.
It is suggested, in this paper, that the ligand shift can give contribution in the constant of the symmetric Heisenberg interaction in antiferromagnetic or ferrimagnetic materials.
Hence, the constant of the Heisenberg interaction is composed minimum of two terms.
One does not depend on the ligand shift an gives standard contribution in the energy density like term with no derivatives of the spin densities or term containing two spatial derivatives of the spin densities.
It is demonstrated that additional term gives a term in the energy density containing one spatial derivative of the spin density.
Corresponding contribution in the Landau--Lifshitz--Gilbert equation is found.
Possibility of the noncollinear equilibrium order of spin under influence of new spin torque is discussed.
Modification of the spin wave (normal modes) dispersion dependencies in the antiferromagnetic materials is found
for the collinear order and for the cycloidal order of spins.
Effective spin current is derived and applied for the spin-current model of the polarization origin in multiferroics.
\end{abstract}

%\pacs{}%
%\keywords{}

\maketitle

%%%%%%%%%%TEXT

%\mbox{\boldmath $\sigma$}

\section{Introduction}

Beginning of the XXI century demonstrated the growth of the interest to the magnetoelectric phenomenon
in the relation to the appearance of the multiferroic materials with the relatively large temperature of the phase transition,
it leads to possible technological applications.
Consequently, understanding of physical mechanisms of the magnetoelectric phenomenon
became highly important issue of the fundamental part of magnetoelectricity field.
Review \cite{Fiebig JP D 05} summarizes some models for the microscopic mechanisms driving magnetoelectric behavior (see Sec. 2.4).
These mechanisms are also discussed over next 20 years in reviews \cite{Nagaosa JP CM 08}, \cite{Tokura RPP 14}, \cite{Dong AinP 15}, \cite{Mostovoy npj 24}.
They are enriched with the spin-current model \cite{Katsura PRL 05} suggested for one of mechanisms,
where the electric dipole moment is proportional to the vector product of spins.
Its macroscopic analog has been discussed for many decades
\cite{Baryakhtar JETP 83}, \cite{Mostovoy PRL 06}.
Some critical analysis of models of the polarization existed in 2008 is given in
\cite{Moskvin PRB 08}.
The spin-current model is initially applied to one regime of the electric polarization appearance \cite{Tokura RPP 14}, \cite{Katsura PRL 05},
but later it is shown
that it can be applied at least to two regimes \cite{AndreevTrukh JETP 24}, \cite{AndreevTrukh PS 24}).
Reexamination of physical mechanisms behind these regimes, 
developed in Refs. \cite{AndreevTrukh JETP 24} and \cite{AndreevTrukh PS 24}), 
applying the extended spin-current model, 
may level out problems pointed out in Ref. \cite{Moskvin PRB 08}.

The magnon spin current associated with the Heisenberg symmetric exchange interaction can be applied in the spin-current model to describe this mechanism
(as it shown in some details on page 7 equations 15-18 in Ref. \cite{Tokura RPP 14},
it is also demonstrated by different method in Refs. \cite{AndreevTrukh JETP 24} and \cite{AndreevTrukh PS 24}).
It leads to the conclussion.
what if some combinations of interactions forms the noncollinear spin order,
the symmetric exchange interaction is responsible for the further formation of the electric polarization.
Later, it is demonstrated
that the electric dipole moment proportional to the scalar product of spins can be described within the spin-current model
with application of different spin current
\cite{AndreevTrukh JETP 24}, \cite{AndreevTrukh PS 24}.
To this end, one needs to use the spin-current of magnons related
to the Dzyaloshinskii-Moriya interaction with the microscopic Dzyaloshinskii constants chosen in the Keffer form.
It shows a partial physical background of the magnetoelectric effect and the contribution of different spin-spin interaction in this phenomenon.

Main goal of this paper is the suggestion of novel form of the spin-spin interaction related to the odd anisotropy of the symmetric
exchange interaction (OASEI).
Presented odd from of the anisotropy is related to the dependence of the exchange integral
on the projection of the interparticle distance between magnetic ions.
Such dependence is possible due to the account of the ligand shift.
Hence, this form symmetric
exchange integral is inspired by the Keffer form of the Dzyaloshinskii constant $\textbf{D}$,
where $\textbf{D}\sim \textbf{r}_{12}\times \mbox{\boldmath $\delta$}$,
with $\mbox{\boldmath $\delta$}$ is the ligand shift
(the shift of nonmagnetic ions from the local center of mass of charges)
and $\textbf{r}_{12}$ is the distance between magnetic ions,
see for instance \cite{Mostovoy npj 24}, \cite{Khomskii JETP 21}.

The Keffer form of the Dzyaloshinskii constant $\textbf{D}\sim \textbf{r}_{12}\times \mbox{\boldmath $\delta$}$
is one example of the analytical form of Dzyaloshinskii constant.
It is possible to identify four forms of the Dzyaloshinskii constant \cite{Andreev 2025 11}.
Two of them can exist in both ferromagnetic and antiferromagnetic materials \cite{Fishman PRB 19}.
Two other forms exist in antiferromagnetic materials only \cite{Andreev 2025 11}.
Suggested in this paper
the odd anisotropy of the symmetric
exchange interaction
can exist in antiferromagnetic materials only.

Both the Dzyaloshinskii-Moriya interaction with the microscopic Dzyaloshinskii constants chosen in the Keffer form
and the symmetric anisotropic exchange interaction with the odd anisotropy
have microscopic Hamiltonian proportional to the interparticle distance of the magnetic ions $\textbf{r}_{12}$.
It leads to the energy density and spin-torque depending on the first derivative on the space coordinate.
It has important consequence for the dispersion dependence of spin waves.
The Dzyaloshinskii-Moriya interaction gives contribution depending on the projection of the wave vector
\cite{Zakeri PRL 10}, \cite{Moon PRB 13}.
It gives the shift of minimum frequency from the zero wave vector.
It appears that the OASEI gives contribution in the dispersion dependence proportional to the square of characteristic frequency,
so it gives contribution proportional to the square of the projection of the wave vector.

Nevertheless, the OASEI has interesting consequences for the polarization formation in the multiferroics.
It gives the effective spin current of magnons,
and it contributes in the spin-current model revealing novel structure of the electric polarization of spin origin.
Existence of additional type of polarization leads to the novel form of the dielectric response
via corresponding spin-torque in addition to the known contributions
\cite{Risinggard SR 16}, \cite{Andreev 2025 05}, \cite{AndreevTrukh EPL 25} for the "noncollinear" parts of spins,
and \cite{Andreev 2025 10} for the "collinear" parts of spins.

Interesting feature of the dielectric response of the multiferroics is the electromagnon resonance,
which is discovered and studied experimentally
\cite{Pimenov NP 06},
\cite{ShuvaevPimenov EPJB 11},
which is also considered theoretically
\cite{AndreevTrukh EPL 25},
\cite{Aupiais npj QM 18},
\cite{Katsura PRL 07},
\cite{Castro PRB 25}.

Regarding the evolution of the small amplitude spin waves in the multiferroic material,
where the the equilibrium distribution of spin density can form periodic cycloidal structure,
we can point out that
it is possible to exclude the periodic coefficients from the equations
for the small amplitude perturbation using simple identical mathematical transformations \cite{Andreev 2025 10}.
But these transformations lead to the appearance of the wave vector of the equilibrium cycloid in the coefficients.
It is true for majority of interactions usually considered in the physics of magnetic and magnetoelectric phenomena.
Both the collinear equilibrium order and the cycloidal equilibrium order in antiferromagnetic materials shows
the independence of two branches of the dispersion dependence,
which actually can be considered as dynamics of $\delta L_{z}$ and $\delta M_{z}$,
where $z$-axis is the direction of anisotropy axis in uniaxial crystal,
and $\textbf{L}=\textbf{S}_{A}-\textbf{S}_{B}$, $\textbf{M}=\textbf{S}_{A}+\textbf{S}_{B}$,
with $\textbf{S}_{A}$ and $\textbf{S}_{B}$ are the spin densities of the magnetic sublattices.
Particularly, one branch of the dispersion dependence of antiferromagnetic materials gives the contribution in the dielectric response
for the multiferroics with the electric polarization proportional
to the scalar product of spin operators
\cite{Andreev 2025 10}.
In Ref. \cite{Andreev 2025 10}, 
it is shown that
the spin wave
with the gapless spectrum contributes to the permittivity for the easy plane equilibrium with the cycloid equilibrium structure.
So, the dielectric response is shifted from the AFM resonance towards smaller frequencies and has small figure-of-merit of resonance since it is proportional to the relatively small parameter $c_{2}$.
Hence, this behavior can be related to the electromagnon observation \cite{Pimenov NP 06}.
The OASEI leads to interference of $\delta L_{z}$ and $\delta M_{z}$.
Hence, it leads to more complex dielectric response of the multiferroics,
where OASEI has the nonzero value.

This paper is organized as follows.
In Sec. II the symmetric
exchange interaction with the odd anisotropy is presented.
In Sec. III the spin torque for the suggested interaction is shown.
In Sec. IV the force field appearing from the novel interaction is presented.
In Sec. V the energy density describing the symmetric
exchange interaction with the odd anisotropy is given.
In Sec. VI the electric polarization caused by the suggested interaction via the spin-current model is obtained.
Corresponding microscopic electric dipole moment is found as well.
In Sec. VII some elements of the XYZ model are discussed.
In Sec. VIII one of mechanisms of the Dzyaloshinskii-Moriya interaction is described.
In Sec. IX contribution of the novel interaction in the dispersion dependence of the spin waves is calculated.
In Sec. X a brief summary of obtained results is presented.

\section{Keffer-like form of the exchange integral in symmetric Heisenberg Hamiltonian}

The matrix form of the coefficient of the Heisenberg Hamiltonian splits
on the symmetric part corresponding to the Heisenberg Hamiltonian with the scalar coefficient
and the antisymmetric part corresponding to the Dzyaloshinskii-Moriya interaction.
The Keffer form of the Dzyaloshinskii-Moriya vector constant includes the contribution of the ligand shift from the local center of mass of charge of ions.
It shows that it is possible to assume the contribution of the ligand shift to the coefficient in the symmetric Heisenberg Hamiltonian.
Obviously this construction includes the requirement of the symmetry of the coefficient $U_{ij}=U_{ji}$.
We start our analysis with the presentation of Heisenberg Hamiltonian with the scalar coefficient
\begin{equation}\label{Ham HHI}
\hat{H}=-\frac{1}{2}\sum_{i,j,j\neq i}U_{ij}(\hat{\textbf{S}}_{i}\cdot \hat{\textbf{S}}_{j}),
\end{equation}
where function $U_{ij}=U(r_{ij})$ is the exchange integral.

\begin{figure}\includegraphics[width=8cm,angle=0]{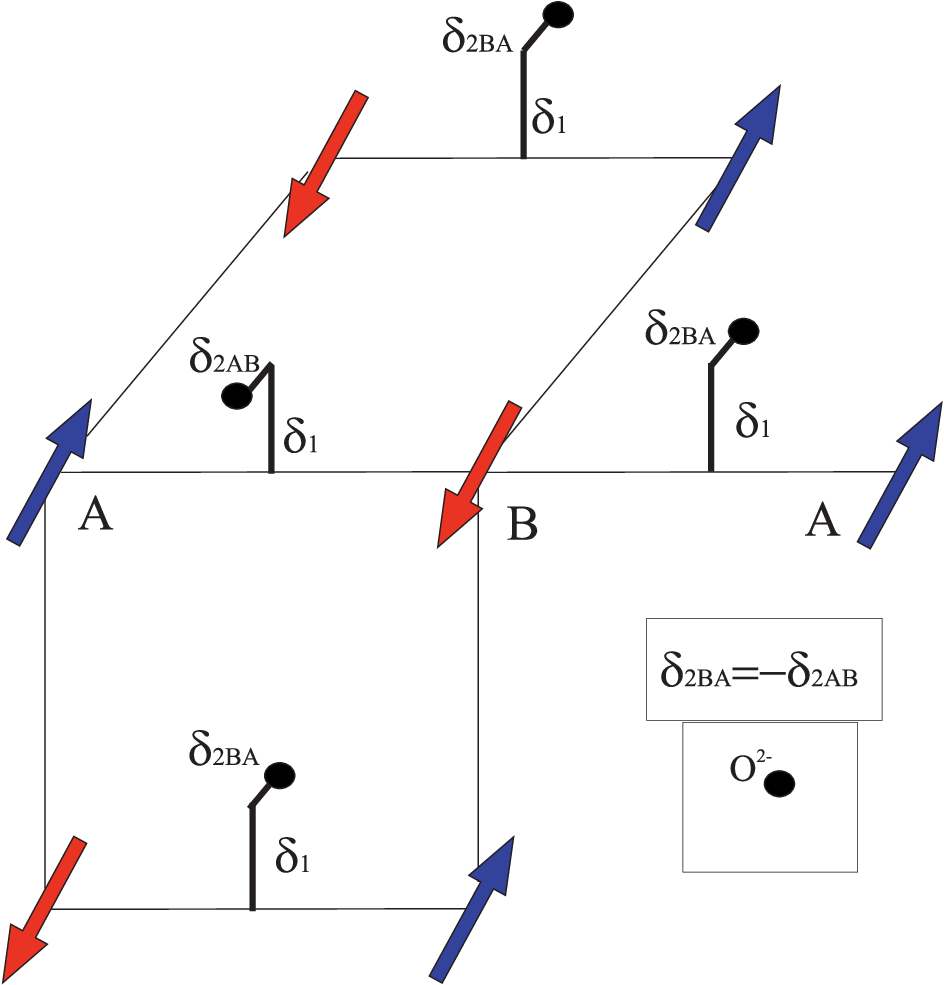}
\caption{\label{Fig 01} The figure shows a possible configuration of the magnetic ions
(demonstrated as arrows) and ligands (oxygen ions demonstrated as black circles) in the antiferromagnetic material.
Partial shifts of the ligand are demonstrated in the figure as
$\mbox{\boldmath $\delta$}_{1}$ and $\mbox{\boldmath $\delta$}_{2,ij-AB}$. }
\end{figure}

\begin{figure}\includegraphics[width=8cm,angle=0]{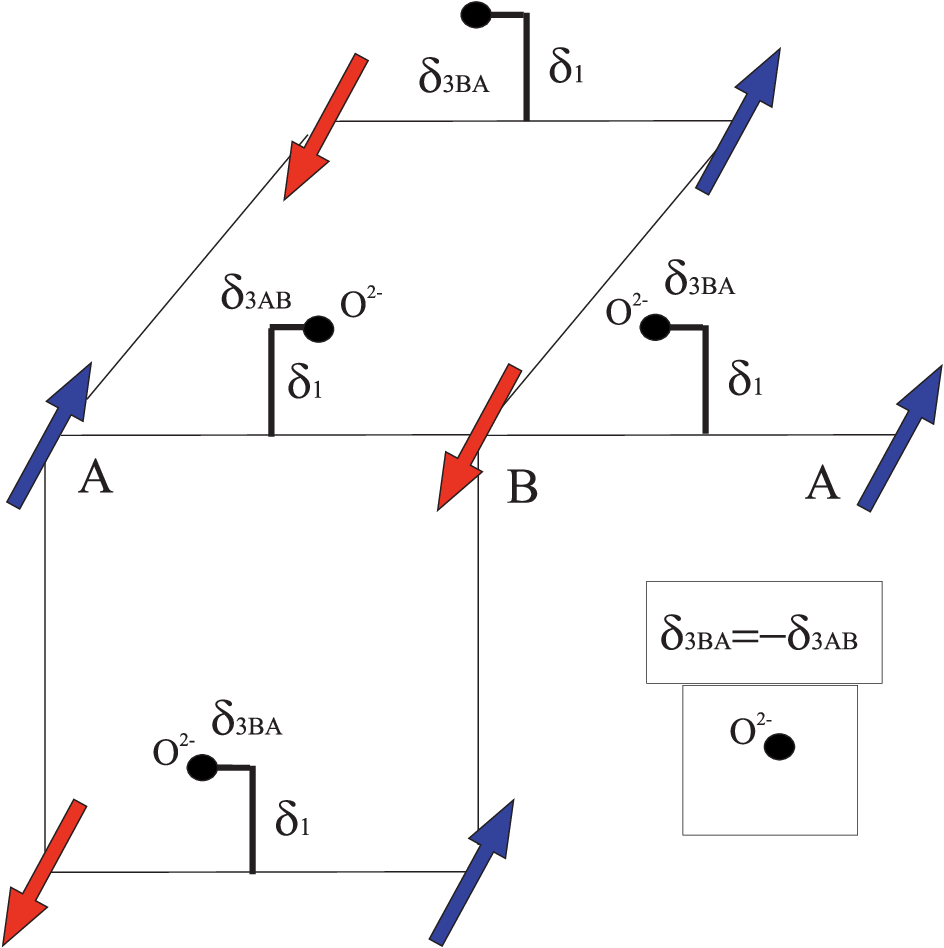}
\caption{\label{Fig 02} This figure is similar to Fig. (\ref{Fig 01}),
but we present the partial ligand shift
$\mbox{\boldmath $\delta$}_{3,ij-AB}$
(which is parallel to the vector connecting magnetic ions $A$ and $B$ $\textbf{r}_{AB}$)
instead of the partial ligand shift $\mbox{\boldmath $\delta$}_{2,ij-AB}$
(which is perpendicular to the plane containing vectors $\textbf{r}_{AB}$ and $\mbox{\boldmath $\delta$}_{1}$).}
\end{figure}

We suggest the following structure for the scalar coefficient (exchange integral) for the antiferromagnetic multiferroic materials
$$U_{ij}=U_{0,ij}+U_{1,ij}(\textbf{r}_{ij}\cdot\mbox{\boldmath $\delta$}_{3,ij-AB})$$
\begin{equation}\label{U structure in Ham HHI}
+U_{2,ij}(\textbf{r}_{ij}\cdot[\mbox{\boldmath $\delta$}_{2,ij-AB}\times\mbox{\boldmath $\delta$}_{1}]),
\end{equation}
where $U_{0,ij}$ is the usual exchange integral.
The second and third terms in equation (\ref{U structure in Ham HHI}) are the suggested odd anisotropic part in the symmetric Heisenberg Hamiltonian.
They includes the partial ligand shifts
$\mbox{\boldmath $\delta$}_{1}$, $\mbox{\boldmath $\delta$}_{2,ij-AB}$, and $\mbox{\boldmath $\delta$}_{3,ij-AB}$
which have the following properties:
$\mbox{\boldmath $\delta$}_{1}$ is a constant,
$\mbox{\boldmath $\delta$}_{2,ji-BA}=-\mbox{\boldmath $\delta$}_{2,ij-AB}$,
and $\mbox{\boldmath $\delta$}_{3,ji-BA}=-\mbox{\boldmath $\delta$}_{3,ij-AB}$.
Described properties allows to ensure the fundamental property of symmetry of the exchange integral $U_{ij}=U(r_{ij})$.
The partial ligand shifts are illustrated in Figs. (\ref{Fig 01}) and (\ref{Fig 02}).

Equation (\ref{U structure in Ham HHI}) is the major concept of this paper.
A number of fundamental consequences are presented below in form of the corresponding macroscopic models
and influence of the suggested term on the collective excitations.
Let us to point out here
that the analysis of the noncollinear equilibrium state allows to estimate the numerical value of $U_{1,ij}$ and $U_{2,ij}$ in terms of known parameters.

The anisotropy of exchange interaction as the tensor structure of generalized exchange integral
$J^{\alpha\beta}_{ij}S_{i}^{\alpha}S_{j}^{\beta}$
is considered in Refs.
\cite{Hermenau Nat Com 19}, \cite{Choi RMP}, \cite{Udvardi PRB 03} (see text after equation 48),
where authors refer to it as the symmetric anisotropic exchange.
It is assumed to be small in comparison with the on-site anisotropy energy.
But the nature of the on-site anisotropy energy is not specified.
Below, considering XYZ-model we obtain macroscopic on-site-like anisotropy energy from the microscopic symmetric anisotropic exchange.
Moreover, we consider the anisotropy related to the functional dependence of the exchange integral as a function of the interparticle distance
(between magnetic ions).
Described odd dependence on the distance appears in addition to the even dependence $J=J(\mid z\mid, \sqrt{x^2+y^2})$.
The last one gives obvious contribution to the anisotropy of the macroscopic exchange term proportional to the second derivatives.

\subsection{Heisenberg constant with combination of two ligand shifts}

The exchange integral
(\ref{U structure in Ham HHI})
includes two additional terms describing the suggested odd anisotropy of the symmetric exchange.
But they have different mechanism.
They can exist in different materials,
so we consider each of them independently.

First, we consider the odd anisotropy of the symmetric exchange,
which can appear in the materials,
where the Dzyaloshinskii-Moriya interaction exists in form of energy density
$\mathcal{E}=-2g_{(0\zeta_{1})}\mbox{\boldmath $\delta$}_{2,AB} [\textbf{S}_{A}\times\textbf{S}_{B}]$.
Both of them required same configuration of the ligand shifts illustrated in Fig. (\ref{Fig 01}).
Here, the Dzyaloshinskii-Moriya energy density contains one partial ligand shift,
which presents the direction of deflection of the ligand from plane formed
by vectors of the relative distance of the magnetic ions $\textbf{r}_{ij}$
and fixed for all ligands shifts $\mbox{\boldmath $\delta$}_{1}$.
Hence, vector $\mbox{\boldmath $\delta$}_{2,AB}$ is perpendicular to the plane of the figure.
This exchange integral explicitly contains two partial ligand shifts
$\mbox{\boldmath $\delta$}_{1}$ and $\mbox{\boldmath $\delta$}_{2,AB}$:
\begin{equation}\label{U structure in Ham HHI 2}
J_{2,ij}=U_{2,ij}(\textbf{r}_{ij}\cdot[\mbox{\boldmath $\delta$}_{1}\times \mbox{\boldmath $\delta$}_{2,ij-AB}]).
\end{equation}
The vector product of the partial ligand shifts $[\mbox{\boldmath $\delta$}_{1}\times \mbox{\boldmath $\delta$}_{2,ij-AB}]$
is parallel to the relative distance of the magnetic ions $\textbf{r}_{ij}$
(see the direction of the partial shifts in Fig. (\ref{Fig 01})).
It gives possible nonzero additional part of the exchange integral $J_{2,ij}$.
We also obtain $J_{2,ij}=J_{2,ji}$ as it is described above (after equation (\ref{U structure in Ham HHI})).

\subsection{Heisenberg constant with the ligand shift parallel to the relative distance}

Another form of the odd anisotropy of the symmetric exchange can be related to the single partial ligand shift
$\mbox{\boldmath $\delta$}_{3,ij-AB}$:
\begin{equation}\label{U structure in Ham HHI 1}
J_{1,ij}=U_{1,ij}(\textbf{r}_{ij}\cdot\mbox{\boldmath $\delta$}_{3,ij-AB}).
\end{equation}
Required partial shift $\mbox{\boldmath $\delta$}_{3,ij-AB}$
exists in addition to $\mbox{\boldmath $\delta$}_{1}$,
both of them are demonstrated in
Fig. (\ref{Fig 02}).

\subsection{Combined contribution of the ligand shifts}

We can present the potential under consideration in a form including both mechanisms
\begin{equation}\label{U structure in Ham HHI eff}
U_{ij}=l_{ij}\delta_{eff,ij-AB}^{\beta}r_{ij}^{\beta},
\end{equation}
where $l_{ij}=U_{1,ij}$ or $l_{ij}=U_{1,ij}$,
and
$\mbox{\boldmath $\delta$}_{eff,ij-AB}=\mbox{\boldmath $\delta$}_{3,ij-AB}$
or
$\mbox{\boldmath $\delta$}_{eff,ij-AB}=[\mbox{\boldmath $\delta$}_{1}\times \mbox{\boldmath $\delta$}_{2,ij-AB}]$,
correspondingly.

\section{Spin torque}

We apply the quantum hydrodynamic method for the derivation of the spin evolution equation
\cite{MaksimovTMP 2001},
\cite{Andreev PoF 21}.
Particularly, we obtain the spin torque corresponding to the Heisenberg Hamiltonian
with the "exchange integral" of form of (\ref{U structure in Ham HHI eff}).
General relation of the spin torque and the Hamiltonian has the following form
\cite{MaksimovTMP 2001},
\cite{Andreev PoF 21}:
\begin{equation}\label{spin time der A}
\partial_{t}\textbf{S}_{A}=\frac{\imath}{\hbar}\int dR\sum_{i\in A}
\delta(\textbf{r}-\textbf{r}_{i})
\Psi^{\dag}(R,t)[\hat{H},\hat{\textbf{S}}_{i}]\Psi(R,t).
\end{equation}
Following results are obtained in an approximate form following
Ref. \cite{AndreevTrukh PS 24}.

Straightforward calculations lead to the following partial contribution of the suggested interaction
(\ref{U structure in Ham HHI eff})
in the Landau--Lifshitz--Gilbert equation for subspecies $A$
\begin{equation}\label{S A evol Ham HHI eff}
\partial_{t}\textbf{S}_{A}=-\frac{1}{3}g_{(l)}[\textbf{S}_{A}\times(\mbox{\boldmath $\delta$}_{eff}\cdot\nabla)\textbf{S}_{B}]
\end{equation}
and, similarly for subspecies $B$
\begin{equation}\label{S B evol Ham HHI eff}
\partial_{t}\textbf{S}_{B}=\frac{1}{3}g_{(l)}[\textbf{S}_{B}\times(\mbox{\boldmath $\delta$}_{eff}\cdot\nabla)\textbf{S}_{A}],
\end{equation}
where
$g_{(l)}\equiv\int r^{2}l(r)d^{3}r$.

It also allows to make transition to the antiferromagnetic vectors
$\textbf{L}=\textbf{S}_{A}-\textbf{S}_{A}$,
and
$\textbf{M}=\textbf{S}_{A}+\textbf{S}_{A}$:
$$\partial_{t}\textbf{L}=\frac{1}{6}g_{(l)}\biggl([\textbf{M}\times(\mbox{\boldmath $\delta$}_{eff}\cdot\nabla)\textbf{M}]$$
\begin{equation}\label{L evol Ham HHI eff}
-[\textbf{L}\times(\mbox{\boldmath $\delta$}_{eff}\cdot\nabla)\textbf{L}]\biggr),
\end{equation}
and
$$\partial_{t}\textbf{M}=\frac{1}{6}g_{(l)}\biggl([\textbf{L}\times(\mbox{\boldmath $\delta$}_{eff}\cdot\nabla)\textbf{M}]$$
\begin{equation}\label{M evol Ham HHI eff}
-[\textbf{M}\times(\mbox{\boldmath $\delta$}_{eff}\cdot\nabla)\textbf{L}]\biggr).
\end{equation}

\section{Force field in the Euler momentum balance equation}

The force field in the Euler equation showing the momentum balance in the physical system corresponding to Hamiltonian (\ref{Ham HHI})
is derived in this section.
The Euler equation plays an important role in the derivation of the spin-current model \cite{AndreevTrukh JETP 24}, \cite{AndreevTrukh PS 24}.
It can be important to understand the complete force field for the analysis of the spin and polarization static and dynamic properties.

The quantum hydrodynamic method \cite{MaksimovTMP 2001}, \cite{Andreev LP 21 fermions}
allows to obtain the following time derivative of the momentum density $m\textbf{j}_{A}=mn\textbf{v}_{A}$
(with the mass of ions $m$, concentration of ions $n$, and the velocity field $\textbf{v}$)
$$\partial_{t}\textbf{j}_{A}=\frac{1}{2m}\frac{\imath}{\hbar}\int dR\sum_{i\in A}
\delta(\textbf{r}-\textbf{r}_{i})\cdot$$
\begin{equation}\label{spin time der A}
\cdot\biggl(\Psi^{\dag}(R,t)[\hat{H},\hat{\textbf{p}}_{i}]\Psi(R,t) -h.c.\biggr),
\end{equation}
with
$[\hat{H},\hat{\textbf{p}}_{i}]=-\imath\hbar\sum_{j\in B}(\hat{\textbf{S}}_{i}\cdot \hat{\textbf{S}}_{j})\nabla_{i} U_{ij}$.

The zeroth order expansion of the delta function $\delta(\textbf{r}-\textbf{r}_{i})$
and wave functions $\Psi(R,t)=\Psi(..,\textbf{r}_{i}, ..., \textbf{r}_{j}, ... ,t)$
on the small relative distance $r_{ij}=\mid \textbf{r}_{i}-\textbf{r}_{j}\mid$
leads to a term containing the following interaction constant
$\int (\delta^{\alpha\beta}l(r)+(r^{\alpha}r^{\beta}/r)(dl(r)/dr))d^{3}r$.
the integration over the angles (and reestablishing the integral over the angles)
simplifies this expression to
$\delta^{\alpha\beta}\int (l(r)+(r/3)(dl(r)/dr))d^{3}r$.
Further integration by part in the second term shows that all interaction constant is equal to zero.

Nonzero value of the force field appears in the second order expansion on the small relative distance
\begin{equation}\label{a}
\textbf{F}_{A}=m\partial_{t}\textbf{j}_{A}
=-\frac{1}{3}g_{(l)}S_{A}^{\gamma}(\mbox{\boldmath $\delta$}_{eff}\cdot\nabla)\nabla S_{B}^{\gamma}.
\end{equation}
Similarly, we derive the force field for the second subspecies $B$:
\begin{equation}\label{a}
\textbf{F}_{B}=m\partial_{t}\textbf{j}_{B}
=\frac{1}{3}g_{(l)}S_{B}^{\gamma}(\mbox{\boldmath $\delta$}_{eff}\cdot\nabla)\nabla S_{A}^{\gamma}.
\end{equation}
These equations can be also represented in terms of the antiferromagnetic vectors
$\textbf{L}=\textbf{S}_{A}-\textbf{S}_{A}$,
and
$\textbf{M}=\textbf{S}_{A}+\textbf{S}_{A}$,
but we do not consider it here.

\section{Energy density}

The spin evolution equation
(the contribution of new interaction in the Landau--Lifshitz--Gilbert equation)
is derived above directly from Hamiltonian (\ref{Ham HHI}) using the many-particle quantum hydrodynamic method \cite{MaksimovTMP 2001}, \cite{Andreev PoF 21}.
The force field is also derived in previous section from Hamiltonian (\ref{Ham HHI}).
Here we obtain the macroscopic energy density.
We do not need it for the further derivation of some equations, since necessary equations are derived above.
But, we need explicit form of the energy density for comprehensive comparison with other known contributions in the model of the magnetically ordered materials, including the comparison with the
Dzyaloshinskii-Moriya interaction.
It is more useful to make comparison with the energy density since it is commonly applied in the study of magnetic materials.
For a start, we present the definition of the macroscopic energy density in terms of the microscopic parameters:
$$\mathcal{E}_{A}=-\frac{1}{2}
\int dR\sum_{i\in A, j\in B}
\delta(\textbf{r}-\textbf{r}_{i})\times$$
\begin{equation}\label{energy definition}
\times\Psi^{\dag}(R,t)U_{ij}(\hat{\textbf{S}}_{i}\cdot \hat{\textbf{S}}_{j})\Psi(R,t).
\end{equation}

Non-zero value of the energy density appears in the first order expansion on the relative distance.
In accordance with the definition,
we find the partial energy density for species "A":
\begin{equation}\label{a}
\mathcal{E}_{A}=\frac{1}{6}g_{(l)}(\textbf{S}_{A}\cdot(\mbox{\boldmath $\delta$}_{eff}\cdot\nabla)\textbf{S}_{B}.
\end{equation}
Similarly, we obtain  the partial energy density for species "B":
\begin{equation}\label{a}
\mathcal{E}_{B}=-\frac{1}{6}g_{(l)}(\textbf{S}_{B}\cdot(\mbox{\boldmath $\delta$}_{eff}\cdot\nabla)\textbf{S}_{A}.
\end{equation}
These partial energy densities can be combined in the full energy density
$\mathcal{E}=\mathcal{E}_{A}+\mathcal{E}_{B}$,
which can be also represented in terms of
the antiferromagnetic vectors
$\textbf{L}=\textbf{S}_{A}-\textbf{S}_{A}$,
and
$\textbf{M}=\textbf{S}_{A}+\textbf{S}_{A}$:   %p.10
\begin{equation}\label{a}
\mathcal{E}=\frac{1}{12}g_{(l)}
\biggl((\textbf{L}\cdot(\mbox{\boldmath $\delta$}_{eff}\cdot\nabla)\textbf{M}
-(\textbf{M}\cdot(\mbox{\boldmath $\delta$}_{eff}\cdot\nabla)\textbf{L}\biggr).
\end{equation}

\section{Spin-current model}

Relation between the electric polarization of spin origin and the spin current is point out in Ref. \cite{Katsura PRL 05} for the case of the electric dipole moment proportional to the vector product of spins
$\textbf{d}_{ij}\sim [\textbf{r}_{ij}\times [\textbf{S}_{i}\times\textbf{S}_{j}]]$, regime of "non-collinear spins".
It corresponds to the relation of polarization to the antisymmetric part of spin current
$P^{\mu}
\sim\varepsilon^{\mu\alpha\beta}J^{\alpha\beta}$.
Estimation of the coefficient in this relation can be found in review \cite{Tokura RPP 14},
where it is related to the matrix element of the electric dipole moment between $p$ and $d$ states.
Later, it is demonstrated that the spin current model can be derived purely in terms of the quantum hydrodynamic model \cite{AndreevTrukh JETP 24} and \cite{AndreevTrukh PS 24}.
So, the macroscopic polarization appears as
\begin{equation}\label{spin current model P} P_{A}^{\mu}
=\frac{\gamma}{c}\varepsilon^{\mu\alpha\beta}J_{A}^{\alpha\beta}.
\end{equation}

Moreover, Refs. \cite{AndreevTrukh JETP 24} and \cite{AndreevTrukh PS 24} show
that two known forms of polarization \cite{Tokura RPP 14}, \cite{Dong AinP 15} %, \cite{}
can be described within the spin-current if corresponding spin currents are chosen.
The effective spin current of magnons,
described by the exchange interaction spin torque in the Landau--Lifshitz--Gilbert equation,
gives the polarization for the regime of "non-collinear spins".
While, the part of spin current of magnons described by the Dzyaloshinskii-Moriya interaction spin torque
(it is the part described by the Keffer form of Dzyaloshinskii constant in the microscopic Hamiltonian)
gives the polarization for
regime of "collinear spins"
$\textbf{d}_{ij}\sim (\textbf{S}_{i}\cdot\textbf{S}_{j})$.
It leads to the general conclusion that the antisymmetric part of the spin current related to some interaction giving a spin torque can lead to an electric polarization.
However, the torques caused by some interactions cannot be presented as the divergence of the effective spin current or this spin current tensor can be symmetric,
so they do not give any contribution in the polarization formation.
In this paper, we suggest the existence of the interaction described by symmetric Heisenberg Hamiltonian with the odd anisotropy.
We assume that it leads to the novel mechanism of the electric polarization formation.

Some critical view on the spin-current model suggested in \cite{Katsura PRL 05}
can be found in Ref.
\cite{Moskvin PRB 08}.
Here we deal with some reexamined spin-current model contracted in \cite{AndreevTrukh JETP 24} and \cite{AndreevTrukh PS 24},
which is based on the quantum hydrodynamic method,
which matches the macroscopic and the microscopic properties of physical systems.

\subsection{Contribution of novel interaction in the polarization formation via the spin-current model}

Complete polarization appears as the sum of the partial polarizations given by the spin-current model
(\ref{spin current model P})
(see also \cite{AndreevTrukh PS 24}).
Hence, the complete polarization is proportional to the sum of the partial spin currents
\begin{equation}\label{a}
J_{\Sigma}^{\alpha\beta}=J_{A}^{\alpha\beta}+J_{B}^{\alpha\beta}
=-\frac{1}{3}\varepsilon^{\alpha\mu\nu}g_{(l_{eff})}\delta_{eff}^{\beta}S_{A}^{\mu}S_{B}^{\nu}.
\end{equation}
It can be considered as the result of the extraction of the the divergence from the complete spin torque $T^{\alpha}$:
$T^{\alpha}=\partial^{\beta}J_{\Sigma}^{\alpha\beta}$,
which is the sum of
(\ref{S A evol Ham HHI eff}) and (\ref{S B evol Ham HHI eff}).
Finally, it gives the polarization
\begin{equation}\label{Pol nm}
\textbf{P}=\frac{1}{3}\frac{\gamma}{c}g_{(l)}
[\textbf{S}_{A}(\textbf{S}_{B}\cdot\mbox{\boldmath $\delta$}_{eff})
-\textbf{S}_{B}(\textbf{S}_{A}\cdot\mbox{\boldmath $\delta$}_{eff})],
\end{equation}
which shows some similarity to the structure suggested in
\cite{Mostovoy PRL 06},
but here we have no space derivatives of the spin density.
Equation (\ref{Pol nm}) shows
the change of the deviation of the spin projections on the effective ligand direction $\mbox{\boldmath $\delta$}_{eff}$.
In more details, we can call $(\textbf{S}_{A}\cdot\mbox{\boldmath $\delta$}_{eff})$ as the local spin projection on the effective ligand direction.
Combination $\textbf{S}_{A}(\textbf{S}_{B}\cdot\mbox{\boldmath $\delta$}_{eff})
-\textbf{S}_{B}(\textbf{S}_{A}\cdot\mbox{\boldmath $\delta$}_{eff})$, for the fixed position,
gives the deviation of the spin projections on the effective ligand direction.
Polarization as the function of position shows the variation of this quantity in space.

Similarly to the energy density definition (\ref{energy definition})
we can introduce the definition of polarization,
which includes the electric dipole moment operator.
So, we need to present the electric dipole moment leading to the macroscopic polarization (\ref{Pol nm}),
which appears in the following form
\begin{equation}\label{a}
\hat{\textbf{d}}_{ij}=\frac{1}{6}\frac{\gamma}{c}r_{ij}^{2}l(r_{ij})
[\mbox{\boldmath $\delta$}_{eff,ij}\times [\hat{\textbf{S}}_{i}\times \hat{\textbf{S}}_{j}]].
\end{equation}

Polarization (\ref{Pol nm}) shows some distant similarity with the spin dependent p–d hybridization mechanism
(see Fig. 2 and equations 21 and 22 in Ref. \cite{Tokura RPP 14}).
However, the last term in equation 21 in Ref. \cite{Tokura RPP 14} describing the microscopic electric dipole moment contains the relative distance of the magnetic ions, which leads to the space derivatives in the macroscopic representation.
Longitudinal part of the electric dipole moment is presented by equation 22 in Ref. \cite{Tokura RPP 14}.
In Ref. \cite{Tokura RPP 14},
above equation 22,
we find different statement
"the contribution due to the
p–d hybridization, which is nonzero for the partially filled $t_{2g}$
orbitals and does not depend on the relative direction of the two spins".
However, the comparison of the second term in equation 21 and equation in Fig. 2d-2f of Ref. \cite{Tokura RPP 14}
(and also the analysis of the spin-current model presented in Refs. \cite{AndreevTrukh JETP 24} and \cite{AndreevTrukh PS 24})
shows that vectors $\textbf{e}$ are the distance on the bond between two sites
(between two magnetic ions, basically).

\subsection{Comparison with known form of polarization}

Equation (\ref{Pol nm}) following from the suggested interaction has some similarity in structure with polarization
\begin{equation}\label{P by Mostovoy}
\textbf{P}\sim (\textbf{S}\cdot\nabla)\textbf{S}-\textbf{S}(\nabla\cdot\textbf{S})
, \end{equation}
which can be found in Refs. \cite{Mostovoy PRL 06}, \cite{Baryachtar JETP Lett 83}, \cite{Stefanovskii Sov. J. Low Temp. Phys 86}, \cite{Sparavigna PRB 94},
and later works including review \cite{Cheong NM 07}.
As it is demonstrated via the spin current model,
polarization (\ref{P by Mostovoy}) can be found using the spin current of magnons given by the symmetric isotropic exchange interaction
\cite{AndreevTrukh JETP 24}, \cite{AndreevTrukh PS 24}.
While polarization (\ref{Pol nm}) follows from the symmetric anisotropic exchange interaction.

\section{Energy density and spin torque for antiferromagnetic materials in XYZ model}

Sections III, IV, V, and VI shows macroscopic consequences of the suggested interaction (\ref{U structure in Ham HHI})
which is the odd anisotropy of the symmetric Heisenberg Hamiltonian.
Before we consider the collective excitations following the model developed above
we describe other interactions included in our calculations.

Mainly we consider the XYZ model for uniaxial crystal with Hamiltonian
$$\hat{H}=-\frac{1}{2}\sum_{n\in A, k\in B}
\biggl(U_{nk}^{xx}\hat{S}_{n}^{x}\hat{S}_{k}^{x} +U_{nk}^{yy}\hat{S}_{n}^{y}\hat{S}_{k}^{y}+U_{nk}^{zz}\hat{S}_{n}^{z}\hat{S}_{k}^{z}\biggr)$$
\begin{equation}\label{Ham XYZ}
=-\frac{1}{2}\sum_{n\in A, k\in B}
\biggl(U_{nk}(\hat{\textbf{S}}_{n}\cdot\hat{\textbf{S}}_{k})
+\tilde{\kappa}_{nk}\hat{S}_{n}^{z}\hat{S}_{k}^{z}\biggr),
\end{equation}
with isotropic coefficients
$U_{nk}^{xx}=U_{nk}^{yy}=U_{nk}^{zz}$,
where
$U_{nk}=U_{nk}^{xx}=U_{nk}^{yy}$,
and
$U_{nk}^{zz}=U_{nk}+\tilde{\kappa}_{nk}$.

\subsection{Energy density in XYZ model}

Similarly to the definition of the energy density (\ref{energy definition})
we define it for Hamiltonian (\ref{Ham XYZ}).
After required simplification
we obtain the energy density associated with one subspecies $A$:
$$\mathcal{E}_{A}=-\frac{1}{2}g_{0,u,AA}(\textbf{S}_{A}\cdot\textbf{S}_{A})-\frac{1}{2}\kappa_{0,AA}(S_{A}^{z}\cdot S_{A}^{z})$$
$$-\frac{1}{12}g_{2,u,AA}(\textbf{S}_{A}\cdot\triangle\textbf{S}_{A})-\frac{1}{12}\kappa_{2,AA}(S_{A}^{z}\cdot \triangle S_{A}^{z})$$
$$-\frac{1}{2}g_{0,u,AB}(\textbf{S}_{A}\cdot\textbf{S}_{B})-\frac{1}{2}\kappa_{0,AB}(S_{A}^{z}\cdot S_{B}^{z})$$
\begin{equation}\label{en density XYZ A}
-\frac{1}{12}g_{2,u,AB}(\textbf{S}_{A}\cdot\triangle\textbf{S}_{B})-\frac{1}{12}\kappa_{2,AB}(S_{A}^{z}\cdot \triangle S_{B}^{z}).
\end{equation}
Expression for the second subspecies can be found from
(\ref{en density XYZ A})
as
$$\mathcal{E}_{B}=\mathcal{E}_{A}(A\leftrightarrow B).$$
Next, the full energy density can be found
$$\mathcal{E}=\mathcal{E}_{A}+\mathcal{E}_{B}.$$

If we apply the common assumption for the relation between the exchange integrals between subspecies
\begin{equation}\label{a}
U_{AB}=-U_{AA}=-U_{BB}\equiv -U
\end{equation}
we get corresponding simplified full energy density
$$\mathcal{E}=-\frac{1}{2}g_{0,u}(\textbf{L}\cdot\textbf{L})-\frac{1}{2}\kappa_{0}(L^{z}\cdot L^{z})$$
\begin{equation}\label{a}
-\frac{1}{12}g_{2,u}(\textbf{L}\cdot\triangle\textbf{L})-\frac{1}{12}\kappa_{2}(L^{z}\cdot \triangle L^{z}).
\end{equation}
Here, we also redefine some coefficients:
$A\equiv g_{2,u}/6$ for the exchange constant,
and $\kappa\equiv \kappa_{0}$ for the anisotropy constant.

Consider the energy of the system corresponding to the "exchange term"
$$E_{ex}=\int \mathcal{E}_{ex}d^{3}r$$
\begin{equation}\label{a}
=-\frac{1}{2}A\int(\textbf{L}\cdot\triangle\textbf{L})d^{3}r=\frac{1}{2}A\int (\partial^{\beta}\textbf{L}\cdot\partial^{\beta}\textbf{L})d^{3}r,
\end{equation}
so this term can be presented in more familiar form.

\subsection{Spin torque in XYZ model}

Above, in this section, we considered the energy density for the XYZ model in a simple assumption $U_{AB}=-U_{AA}=-U_{BB}\equiv -U$.
But further analysis we make for more general case of $U_{AA}=U_{BB}\neq \mid U_{AB}\mid$.
It leads to the following combinations of the exchange integral in the spin evolution equations
$U_{+}=(U_{AA}+U_{BB}+2 \mid U_{AB}\mid)/4$, and $U_{-}=(U_{AA}+U_{BB}-2 \mid U_{AB}\mid)/4$.
Described combined exchange integrals $U_{+}$ and $U_{-}$ lead to two sets of the coefficients in the spin evolution equations
$$\partial_{t}\textbf{L}=-g_{0u}[\textbf{L}\times \textbf{M}]
+\kappa_{+}[\textbf{M}\times L_{z}\textbf{e}_{z}]+\kappa_{-}[\textbf{L}\times M_{z}\textbf{e}_{z}]$$
$$+A_{+}[\textbf{M}\times \triangle \textbf{L}]+A_{-}[\textbf{L}\times\triangle\textbf{M}]$$
\begin{equation}\label{L evol XYZ AB}
+\frac{1}{6}\kappa_{2+}[\textbf{M}\times \triangle L_{z}\textbf{e}_{z}]
+\frac{1}{6}\kappa_{2-}[\textbf{L}\times\triangle M_{z}\textbf{e}_{z}],
\end{equation}
and
$$\partial_{t}\textbf{M}=
\kappa_{+}[\textbf{L}\times L_{z}\textbf{e}_{z}]+\kappa_{-}[\textbf{M}\times M_{z}\textbf{e}_{z}]$$
$$+A_{+}[\textbf{L}\times \triangle \textbf{L}]+A_{-}[\textbf{M}\times\triangle\textbf{M}]$$
\begin{equation}\label{M evol XYZ AB}
+\frac{1}{6}\kappa_{2+}[\textbf{L}\times \triangle L_{z}\textbf{e}_{z}]
+\frac{1}{6}\kappa_{2-}[\textbf{M}\times\triangle M_{z}\textbf{e}_{z}].
\end{equation}
We apply this set of symmetric Heisenberg Hamiltonian based
the Landau--Lifshitz--Gilbert equations
in combination with the spin torque following from the suggested interaction
(\ref{L evol Ham HHI eff}) and (\ref{M evol Ham HHI eff})
and a part Dzyaloshinskii-Moriya interaction described in the next section.

\section{Dzyaloshinskii-Moriya interaction}

We do not discuss the Dzyaloshinskii-Moriya interaction in details,
but following Refs. \cite{Andreev 2025 11}, \cite{Andreev 2025 Vestn},
we consider a part of the Dzyaloshinskii-Moriya interaction with the Dzyaloshinskii constant of the following form
\begin{equation}\label{a}
\textbf{D}_{ij,(\delta)}=\zeta_{1}(r_{ij})\mbox{\boldmath $\delta$}_{2,ij-AB},
\end{equation}
which leads to the well-known form of the energy density
\begin{equation}\label{DMI D en}
\mathcal{E}=-2g_{(0\zeta_{1})}\mbox{\boldmath $\delta$}_{2,AB} [\textbf{S}_{A}\times\textbf{S}_{B}],
\end{equation}
and corresponding spin torques
\begin{equation}\label{a}
\partial_{t}\textbf{S}_{A,DMI}=
-g_{(0\zeta_{1})}[\textbf{S}_{A}\times [\mbox{\boldmath $\delta$}_{2,AB} \times \textbf{S}_{B}]]
\end{equation}
or for the collective functions
\begin{equation}\label{L evol DMI 2}
\partial_{t}\textbf{L}=\frac{1}{2}g_{(0\zeta_{1})}
\biggl([\textbf{L}\times [\mbox{\boldmath $\delta$}_{2,AB}\times\textbf{L}]]
-[\textbf{M}\times [\mbox{\boldmath $\delta$}_{2,AB}\times\textbf{M}]]\biggr),
\end{equation}
and
\begin{equation}\label{M evol DMI 2}
\partial_{t}\textbf{M}=\frac{1}{2}g_{(0\zeta_{1})}[\mbox{\boldmath $\delta$}_{2,AB}\times[\textbf{M}\times\textbf{L}]].
\end{equation}
Some recent applications of this form of the Dzyaloshinskii-Moriya interaction can be found in Ref. \cite{Gareeva PRB 13}.

We consider this part of the Dzyaloshinskii-Moriya interaction
since it is required to form equilibrium state for the cycloidal spiral equilibrium configuration.
This Dzyaloshinskii-Moriya interaction balance the odd anisotropy of the exchange interaction suggested above.

\section{Spin wave dispersion dependence: contribution of the odd anisotropy of the Heisenberg exchange}

Contribution of the suggested in this paper odd anisotropy of the Heisenberg exchange in the properties of fundamental collective excitations is under consideration in this section.

Dzyaloshinskii-Moriya interaction (\ref{DMI D en}) can lead to nonzero equilibrium value of vector
$\textbf{M}=\textbf{S}_{A}+\textbf{S}_{B}$,
but we consider the equilibrium states with $\textbf{M}_{0}=0$.

\subsection{Collinear order - easy axis regime}

\subsubsection{Equilibrium}

We assume $\textbf{L}_{0}=L_{0}\textbf{e}_{z}$ and $\textbf{M}_{0}=0$,
with $L_{0}=const$.

The spin torques given by equations
(\ref{L evol Ham HHI eff}) and (\ref{M evol Ham HHI eff})
are equal to zero.
Same result is correct for equations
(\ref{L evol XYZ AB}) and (\ref{M evol XYZ AB}).
In these four equations each term is equal to zero for the chosen equilibrium.

Equations (\ref{L evol DMI 2}) and (\ref{M evol DMI 2}) are famous in relation to the a possibility of formation of nonzero $\textbf{M}_{0}$,
but here we consider $\textbf{M}_{0}=0$.
The first term in equation (\ref{L evol DMI 2}) can be equal to zero if $\textbf{L}_{0}\parallel \mbox{\boldmath $\delta$}_{2,AB}$. 
Other terms in equations (\ref{L evol DMI 2}) and (\ref{M evol DMI 2}) are equal to zero for this equilibrium state.

\subsubsection{Perturbations}

The dispersion dependence for the small amplitude of the described equilibrium appears
as
\begin{equation}\label{omega CoEa}
\omega^{2}=\Omega_{2}^{2}+\Omega_{3}^{2}\pm\sqrt{\omega_{0}^{4}+4\Omega_{2}^{2}\Omega_{3}^{2}},
\end{equation}
where
$\Omega_{2}=g_{0\zeta}L_{0}\delta_{2}/2$ is the characteristic frequency of
the Dzyaloshinskii-Moriya interaction,
$\Omega_{3}=g_{(l)}L_{0}(\mbox{\boldmath $\delta$}_{eff}\cdot \textbf{k})/6$
is the characteristic frequency of
the symmetric exchange interaction with the odd anisotropy,
and
\begin{equation}\label{a}
\omega_{0}^{2}=(\kappa +A_{+}k^2)(\kappa +g_{0u} +A_{-}k^2)L_{0}^{2}
\end{equation}
which is the frequency square for the spin waves in the absence of
the Dzyaloshinskii-Moriya interaction with the energy density (\ref{DMI D en}) and
the symmetric exchange interaction with the odd anisotropy.

The dispersion dependence (\ref{omega CoEa}) contains two types of solutions.
Solution with $+$ in front of the square root corresponds to the modification of the spin wave dispersion dependence existing without
the Dzyaloshinskii-Moriya interaction with the energy density (\ref{DMI D en}) and
the symmetric exchange interaction with the odd anisotropy.
Solution with $-$ in front of the square root is negative $\omega^{2}<0$.
It can lead to the instability of the presented equilibrium,
which can be related to the appearance of the nonzero $\textbf{M}_{0}$.

\subsection{Collinear order - easy plane regime}

\subsubsection{Equilibrium}

We assume $\textbf{L}_{0}=L_{0}\textbf{e}_{x}$ and $\textbf{M}_{0}=0$,
with $L_{0}=const$.
These conditions satisfy equations
(\ref{L evol Ham HHI eff}), (\ref{M evol Ham HHI eff}),
(\ref{L evol XYZ AB}), (\ref{M evol XYZ AB}),
and
(\ref{M evol DMI 2}),
where each term is equal to zero.

Equation (\ref{L evol DMI 2}) leads to the following condition for the relative orientation of the equilibrium antiferromagnetic vector and the ligand shift
$\mbox{\boldmath $\delta$}_{2,AB}\parallel \textbf{L}_{0}=L_{0}\textbf{e}_{x}$
(which is opposite to the previous subsection, where $\mbox{\boldmath $\delta$}_{2,AB}$ is parallel to the anisotropy axis).

\subsubsection{Perturbations}

We consider the small amplitude perturbations of the described equilibrium state.
It leads to the following dispersion dependence
$$\omega^{2}=\frac{1}{2}\biggl[\omega_{1}^{2}+\omega_{2}^{2}-2(\Omega_{2}^{2}+\Omega_{3}^{2})$$
\begin{equation}\label{omega on k easy plane}
\pm\sqrt{(\omega_{1}^{2}-\omega_{2}^{2})^2+16\Omega_{2}^{2}\Omega_{3}^{2}- 4 \tilde{\kappa}\tilde{\kappa}_{-}L_{0}^{2}\Omega_{3}^{2}}
\biggr],
\end{equation}
where
$\tilde{\kappa}\equiv\kappa-\frac{1}{6}\kappa_{2+}k^2$,
$\tilde{\kappa}_{-}\equiv\kappa_{-}+\frac{1}{6}\kappa_{2-}k^2$,
\begin{equation}\label{a}
\omega_{1}^{2}=(g_{0u}+A_{-}k^2)\biggl(-\kappa+A_{+}k^2+\frac{1}{6}\kappa_{2+}k^2\biggr)L_{0}^{2},
\end{equation}
and
\begin{equation}\label{a}
\omega_{2}^{2}=\biggl(g_{0u}+\kappa_{-}+A_{-}k^2+\frac{1}{6}\kappa_{2-}k^2\biggr)A_{+}k^2 L_{0}^{2},
\end{equation}
with $\kappa<0$.

Here $\omega_{1}(k)$ and $\omega_{2}(k)$ give the dispersion dependencies of two spin waves existing at the zero contribution
of the Dzyaloshinskii-Moriya interaction with the energy density
(\ref{DMI D en}) and
the symmetric exchange interaction with the odd anisotropy.

Both regimes (easy axis and easy plane) show that the characteristic frequency of the symmetric exchange interaction with the odd anisotropy
$\Omega_{3}=g_{0l}L_{0}(\mbox{\boldmath $\delta$}_{eff}\cdot \textbf{k})/6$
appears in the second degree.
Hence, we find no dependence of the frequency on the direction of the wave propagation. 
It shows the difference with the Dzyaloshinskii-Moriya interaction describe by the Lifshitz invariants \cite{Fishman PRB 19}, 
where there is the dependence on the wave vector projection.

\subsection{Cycloidal order}

\subsubsection{Equilibrium}

We consider the noncollinear equilibrium condition, 
which is also corresponds to the easy-plane regime of the uniaxial crystals
\begin{equation}\label{a}
\textbf{L}_{0}=L_{b}\cos(qx)\textbf{e}_{x} +L_{c}\sin(qx)\textbf{e}_{y},
\end{equation}
and
$\textbf{M}_{0}=0$,
where $L_{c}=\pm L_{b}$.

In this case the first term in equation
(\ref{L evol Ham HHI eff})
has nonzero projection on the $z$-direction.
Corresponding spin torque $T_{oa}^{z}=(1/6)g_{(l)}(\mbox{\boldmath $\delta$}_{eff}\cdot \textbf{q})L_{b}^{2}(L_{b}/L_{c})$.
It is nonzero if $\mbox{\boldmath $\delta$}_{eff}$ has nonzero projection on the wave vector of equilibrium cycloid $\textbf{q}=q\textbf{e}_{x}$.

All terms in equations
(\ref{M evol Ham HHI eff}),
(\ref{L evol XYZ AB}), (\ref{M evol XYZ AB}),
(\ref{M evol DMI 2}) are equal to zero (each term).

However, we have nonzero value of the first term in equation
(\ref{L evol DMI 2}):
$\textbf{T}_{DM}=g_{0\zeta}L_{b}^{2}\mbox{\boldmath $\delta$}_{2}/2$.
Hence, there is the balance at $\mbox{\boldmath $\delta$}_{2}\parallel \textbf{e}_{z}$.

Thus, the balance of these conditions allows to estimate unknown parameter:
the interaction constant of the suggested symmetric exchange interaction with the odd anisotropy $g_{(l)}$.
Hence, we find
\begin{equation}\label{eq g2l and gzeta}
g_{(l)}=\frac{3g_{0\zeta}}{q}\frac{\delta_{2}}{\delta_{eff}}(L_{c}/L_{b}).
\end{equation}
If $\delta_{eff}=\delta_{1}\delta_{2}$,
we obtain $\delta_{2}/\delta_{eff}=1/\delta_{1}$.

Analysis of the dispersion dependencies in the collinear regime allowed to introduce characteristic frequencies
$\Omega_{2}$ and $\Omega_{3}$.
Presence of the cycloid allows to introduce an analog of $\Omega_{3}$:
$\Omega_{4}=(q/k)\Omega_{3}=(1/6)g_{(l)}L_{0}\delta_{eff}q$. 
Equilibrium condition
(\ref{eq g2l and gzeta}) shows $\mid\Omega_{2}\mid=\mid\Omega_{4}\mid$.

\subsubsection{Perturbations}

Analysis of perturbations at the cycloidal equilibrium order can be found in Ref. \cite{Andreev 2025 10}.
Similarly, we find that the presence of the periodic equilibrium order shifts some coefficients in the dispersion dependence via
the additional terms depend on $q$.
While the periodic coefficients are excluded without neglecting any of them.
This conclusion is in general agreement with works \cite{Bychkov JMMM 13} and \cite{Bychkov SSP 12},
where authors considered periodic equilibrium in ferromagnetic materials.
Both the collinear order-easy plane regime
and the cycloidal order-easy plane regime shows independent evolution of $\delta L_{z}$ and $\delta M_{z}$
if $\Omega_{3}$ related to the OASE interaction is negligible.
However, influence of $\delta L_{z}$ and $\delta M_{z}$ on each other happens due to the OASE interaction $\Omega_{3}\neq0$.
It is illustrated via the dispersion equations in the Appendix.

\section{Conclusion}

I has been demonstrated
that the ligand shift usually responsible for the appearance of the Dzyaloshinskii-Moriya interaction
affects the symmetric Heisenberg interaction in antiferromagnetic materials
(or other multicomponent magnetically ordered systems).
It leads to qualitatively new spin torques in the Landau--Lifshitz--Gilbert equation,
and, therefore, modifies the static and dynamical behavior of the magnetically order systems.
This modification is illustrated via the dispersion dependence of spin waves,
which has been derived in this paper.
New from of the polarization of spin origin appears from the suggested interaction via corresponding spin torque and the generalized spin-current mechanism.
These examples show a possibility of various application of suggested interaction to antiferromagnetic and multiferroic materials.

\section{DATA AVAILABILITY}

Data sharing is not applicable to this article as no new data were
created or analyzed in this study, which is a purely theoretical one.

\section{Acknowledgements}

The work is supported by the Russian Science Foundation under the
grant
No. 25-22-00064.

\appendix

\section{On structure of the dispersion equations in easy plane regime}

For the collinear order, 
we obtain
the following dispersion equation
$$(\imath\omega+\Omega_{2})^2\delta M_{z}=\Omega_{3}^2 \delta M_{z}
+A_{+}L_{0}k^2\frac{\imath\omega+\Omega_{2}}{-\imath\omega+\Omega_{2}}\times$$
$$\times
\biggl([g_{0u}-\kappa_{0-}+A_{-}k^2+\frac{1}{6}\kappa_{2-}k^2]L_{0}\delta M_{z}-\imath\Omega_{3}\delta L_{z}\biggr)$$
\begin{equation}\label{a} +\imath\Omega_{3}[\kappa-A_{+}k^2-\frac{1}{6}\kappa_{2+}k^2]L_{0}\delta L_{z},\end{equation}
where $\kappa<0$,
and
$$(-\imath\omega+\Omega_{2})^2\delta L_{z}=\Omega_{3}^2 \delta L_{z}
+(g_{0u}+A_{-}k^2)L_{0} \frac{-\imath\omega+\Omega_{2}}{\imath\omega+\Omega_{2}}\times$$
$$\times
\biggl(-L_{0}[\kappa-A_{+}k^2-\frac{1}{6}\kappa_{2+}k^2] \delta L_{z}
+\imath\Omega_{3}\delta M_{z}\biggr)$$

\begin{equation}\label{a}
+\imath\Omega_{3} [g_{0u}-\kappa_{0-}+A_{-}k^2+\frac{1}{6}\kappa_{2-}k^2]L_{0}\delta M_{z}.\end{equation}
It shows that the existence of the OASE interaction (presented by $\Omega_{3}$) creates the coupling between these equations, 
and corresponding spin waves.

\end{document}